\documentclass[12pt]{article}
\usepackage[english]{babel}
\usepackage{amsmath,amssymb}
\usepackage{url}
\usepackage{xcolor}
\usepackage[authoryear,round]{natbib}
\usepackage{rotating}
\usepackage[margin=20mm]{geometry}
\usepackage{hyperref}
\hypersetup{
    colorlinks = true,
    linkcolor = blue,
    anchorcolor = blue,
    citecolor = blue,
    filecolor = blue,
    urlcolor = blue,
	pdfauthor={Luigi Foschini},
	pdftitle={Matter without mass}
    }

\begin{document}

\title{Matter without mass:\\do we really need the concept of mass?}
\author{Luigi Foschini\\
\small{Brera Astronomical Observatory}\\
\small{National Institute of Astrophysics (INAF)}\\
\small{Milano/Merate, Italy}\\
\small{Email: \texttt{luigi.foschini@inaf.it}}}
\date{November 18, 2026}
\maketitle

\begin{abstract}
Einstein's most famous equation -- $E=mc^2$ -- generated a short-circuit between the concepts of mass and energy, which also affects other concepts like matter, radiation, and vacuum. Physics currently has a mixture of classical, relativistic, and quantum concepts of mass, which generates a great deal of confusion and many problems. Clear definitions need to be established if one wants to avoid ghost hunting. In particular, by abandoning the idea of mass as a primitive concept and focusing on time and energy, some interesting implications are emerging. It is noted that the cutoff frequency of the quantum vacuum energy, which would reproduce the observed value of the cosmological constant, is within one order of magnitude of the present-day peak frequency of the Cosmic Microwave Background. The previous versions of this essay speculated on a physical link between the two frequencies, and on a rotating universe described by a Kerr-de Sitter metric. In this revised version, I show that the coincidence holds only at the present epoch -- it is another way to state the well-known cosmic coincidence problem -- and that the speculations built on it must be dropped.\\

\end{abstract}

\section{Introduction}
Mass is still one of the fundamental concepts of physics. Its definition developed and changed over the centuries, from Newton (quantity of matter in a given volume) to Einstein (measure of the energy content of a physical body), with random outbursts of debates such as the renormalization in quantum electrodynamics, the relativistic mass, and the genesis of the mass via the Higgs boson (see \citealt{BAGGOTT,JAMMER} for detailed and pleasant reviews). These debates were never truly solved, but simply bypassed. Yet, it is still possible to see different mass concepts mixed together in different research fields. Perhaps, it is time to face this problem again and to adopt a radical solution: the removal of mass as a primitive concept, at least from frontier physics. There is no need to abandon it in classical physics, where it still makes sense. 

The idea is not new. \cite{WHEELER} coined the expression ``mass without mass''\footnote{I think this expression is misleading, even though it has been taken up more recently by \cite{WILCZEK02,WILCZEK12}. I prefer to speak of matter without mass, as I have written in the title of this essay.} for the geon, a standing electromagnetic wave held together by its own gravitational attraction: seen from a distance, it attracts and moves like any other mass, yet there is no place inside it where one can say ``here is mass''. The geon is a classical and unstable object, and Wheeler himself stressed that it is not meant to explain anything observed, but it shows that general relativity does not need mass as a primitive concept. \cite{WILCZEK02,WILCZEK12} revived the expression for quantum chromodynamics: about 95\% of the mass of protons and neutrons, and thus of ordinary matter, is the energy of the quark and gluon fields in a theory whose equations contain almost no mass parameters. Another attempt to explain rest-mass energy as a gravitational effect has been proposed by \cite{MELIA}: in a universe whose apparent horizon expands at the speed of light, the energy $mc^2$ can be identified with the gravitational binding energy of a particle to the Misner-Sharp mass within that horizon. The derivation relies, however, on the same equivalence $m=E/c^2$ in the definition of the gravitating mass, and applies only under a specific equation of state not satisfied by the standard $\Lambda$CDM model \citep{MELIA}. 

Following these authors, what is proposed here is not to deny that bodies have a rest energy, or that $m=E/c^2$ is a well-defined, Lorentz-invariant parameter labelling them, but to stop treating mass as a primary quality of matter. In modern physics, energy and momentum are the primary dynamical concepts, and the Newtonian mass is an approximate, emergent property of matter. What remains unexplained are the masses of the elementary quanta and, most of all, the origin of dark matter and dark energy. 

At first glance, particle physicists already did it by adopting natural units, so that every quantity is in energy units. This choice is justified, as noted by the French philosopher \cite{BACHELARD}, by the fact that a physical particle is a real Kant's \emph{noumenon}, an unaccessible physical reality, but with hints of it emerging from experiments (see also \citealt{DAVIES1984}). The particle of modern physics is not a microscopic version of a human-sized object, like a ball or a spinning top. The size of a particle is determined in a dynamical way, as it depends on its cross section for collision with other particles. It has no geometrical size, but it has a temporal size determined by its energy. Space derives from the relationships between these particles, which are nothing more than knots of energy. These ideas recall somehow the thought by \cite{MARKOPOULOU}, with time/energy as primary quantities and the space emerging from relationships between objects. 

However, adopting natural units is not enough. Each research field has its own set of natural units, which in turn generates different results. Particle physicists adopt $\hbar=c=1$, while astrophysicists and cosmologists use $G=c=1$, where $\hbar=h/2\pi$ is the reduced Planck constant, $c$ is the speed of light in vacuum, $G$ is the gravitational constant. A mass is expressed in terms of energy for the former, but as a length for the latter. Sometimes, the choice to set some constants -- and not others -- equal to one just depends on the personal taste of the individual researcher. This obviously generates confusion, and, most important, the idea of mass still lingers around. This does not mean that natural units are mere calculation tricks, as I wrote in the previous versions of this essay: there are profound reasons to regard $c$, $\hbar$, and $G$ as conversion factors -- between space and time, between frequency and energy, between energy density and space-time curvature -- rather than as numerical parameters \citep{WILCZEK02}. These constants are also crucial for maintaining a link to experiments and observations.

The two lengths associated with a mass $m$ by the two communities are the Compton wavelength $\hbar/mc$ and the gravitational radius $Gm/c^2$: the former decreases with $m$, the latter increases, and their product is always the square of the Planck length, $G\hbar/c^3$. The very concept of particle is hinged on the concept of mass. Quantum field theory does not remove it: on the contrary, it explains why all the quanta of a given field -- e.g. all the electrons -- have exactly the same mass \citep{WILCZEK12}. But it changes its status, from a primary quality of matter to a parameter of the field and, for composite bodies, to the energy of the fields in the rest frame. Therefore, removing mass as a primitive concept requires also some changes in our way of thinking, particularly to draw more attention on time and energy. 

\section{Relativistic mass}
The well-known equation $E=mc^2$ raised the debate on the so-called relativistic mass, rest plus kinetic, which was a misleading concept, now abandoned \citep{OKUN}: the kinetic term is simply a change of the rhythm of time, since if one uses the proper time to measure velocity, the definition of momentum remains unchanged (e.g. \citealt{RINDLER}).

Focusing on time, and its change of rhythm due to gravitation, should make it easier to understand the peculiar role of the photon. It has no rest mass, but it has momentum and energy. When passing through a gravitational field, which is nothing else than a significant concentration of energy, it simply changes its rhythm of time (red-/blue-shift of frequency) depending on the position in the gravitational field and its direction of motion. An extreme example is the geon: a standard clock inside it ticks at one third of the rate of a standard clock far away \citep{WHEELER}. The photon, as a massless particle, can be thought as a universal clock, if one is ready to abandon the assumption of uniformity and constancy of the rhythm of time. Since the photon has no rest energy, there is no rest frame, and then no proper or comoving time. William Unruh wrote that ``gravity \emph{is} the unequable flow of time from place to place'' rather than being the cause of a change of the rhythm of time \citep{UNRUH}. However, this does not imply that time and energy are the same: energy is what is needed to translate in time, but it is not time.

\section{Rest, inert, Newtonian mass}
In classical physics, there is a boundary between matter and radiation -- a boundary that, as we will see, is less sharp than it appears. The classical definition is clearly based on macroscopic objects. The mass is the quantity of matter (solid, liquid, or gaseous) in a given volume, while the radiation is the electromagnetic field. Everything is straightforward. 

Quantum mechanics showed us that solid, liquid or gaseous physical bodies are composed of molecules and atoms, which in turn are made of quarks and leptons. Electromagnetic and strong nuclear interactions bind these particles. However, both particles and bonds are energy. Once again, if one does not think about particles, but about quantum excitations of fields, it is immediately clear the role of energy. As already noted, about 95\% of the mass of protons and neutrons is the energy of the quark and gluon fields confined inside them \citep{WILCZEK12}.

In classical physics, matter was distinguished from radiation by its rest energy: matter has energy even when at rest in some reference frame, while free radiation does not. The difference is, more precisely, between confined and free energy, not between two different substances. A single photon has no rest frame, but a system of photons generally has one: two photons moving in opposite directions have a non-zero invariant mass, and radiation confined in a region -- by its own gravity in the geon \citep{WHEELER}, or the massless gluon field by the strong interaction in the proton \citep{WILCZEK02,WILCZEK12} -- has a rest energy and behaves as a massive body. A tiny part of this energy is necessary to translate in time, to exist at rest (existence energy\footnote{Curiosity: the word atom derives from Greek \emph{a-tomos}, where the prefix a- stands for not, while \emph{tomè} (cut) derives from \emph{tèmnein}, to cut. An atom would be something that cannot be cut. The word time has the same root, \emph{tem-n\^{o}}, and means to cut (see \citealt{FOSCHINI} for an extended etymological analysis.). Therefore, the atom, the basic constituent of the matter, is not something that cannot be broken -- as commonly thought -- but it is something not subject to the cut of time.}, see \citealt{FOSCHINI}). Most of energy is divided into bonds between particles, either molecular or subatomic bonds, and can be partially released by radioactive decay, or by breaking molecular or nuclear bonds. It is just a problem of different types of energy, condensed in different ways. Therefore, again, the concept of mass as a primary quality of matter is no more necessary. 

To summarize, matter is an aggregate of different types of energy, and one part is spent to exist at rest. Free radiation is pure temporal energy, it is a universal clock, it cannot exist at rest, because it would be in contradiction with the concept of time. It is interesting to note that in the Standard Model, only two bosons are massless, the photon and the gluon. Both are responsible for generating bonds between particles, but the photon is more important because it can be used as a clock, while the very short range of the strong force makes the gluon useless for such a purpose (also the self-interaction of this boson hampers the possibility to serve as clock). 

\section{Dark Matter}
In the 1930s, Fritz Zwicky first speculated the existence of a non-luminous matter by observing the radial velocities of galaxies in clusters. Later, Vera Rubin reported anomalous rotation curves of galaxies. The most common explanation is to search for some exotic and unknown particle (see \citealt{CIRELLI} for a recent review), but despite decades of experiments, no detection was ever reported (e.g. see \citealt{APRILE} for one of the latest experimental reports). On the contrary, the search for other explanations has produced important discoveries. By using the superior astrometry of the \emph{Gaia} satellite, \cite{CROSTA} proposed that the rotation curve of the Milky Way can be explained by the contribution to the gravitational field of an off-diagonal term in the metric (due to rotation), without any need to invoke some exotic matter. The result was confirmed with Gaia DR3 \citep{BEORDO2024A} and extended to MOND and $\Lambda$CDM, all models being statistically equivalent in representing the observed velocities \citep{BEORDO2024B}. The physical interpretation is debated: in the weak-field, low-velocity regime appropriate for disk galaxies, the general relativistic corrections to the Newtonian rotation curve are of the order of $v^2/c^2\sim 10^{-6}$ \citep{CIOTTI}, and the exact solution on which the model is based was argued to be inappropriate as a galactic model \citep{COSTA}. The authors replied to these criticisms \citep{BEORDO2024A}, while acknowledging that the origin of the gravitational dragging remains undetermined. MOND theory was instead successfully adopted to explain the rotation curves of a sample of galaxies \citep{CHAE}.

Be that as it may, the \cite{CROSTA} result is particularly interesting, because one might think of searching for a similar effect (rotation) on cosmological scales. To test this hypothesis, it is necessary to prove the rotation of the Universe. \cite{GODEL} showed that a rotating universe would have an anisotropic distribution of galaxies, with an excess in one hemisphere. This must not be confused with the dipole anisotropy of the Cosmic Microwave Background (CMB) due to the motion of the Solar System, the Milky Way, and the Local Group \citep{STEWART}, nor with the corresponding dipole expected in the number counts of distant sources \citep{ELLIS}, whose observed amplitude is anomalously large (see \citealt{SECREST25} for a review). However, a global rotation is already severely constrained by the isotropy of the CMB: the analysis of the \emph{Planck} temperature and polarization maps in the framework of Bianchi cosmologies resulted in an upper limit for the vector mode (associated with vorticity) of $(\sigma_{V}/H)_{0}<4.7\times 10^{-11}$ \citep{SAADEH}. This is consistent with the expectation that any primordial rotation is exponentially suppressed during an inflationary epoch, which makes the smallness of the present-day vorticity a natural consequence of the early expansion \citep{ELLISO}. Therefore, any rotation of the universe, if present, must be extremely small. 

Another interesting result regards angular momentum on very large scales: \cite{WANG} stacked $\approx 17000$ filaments of galaxies from the \emph{Sloan Digital Sky Survey} and found a redshift difference between the galaxies on the two sides of the filament spine, consistent with a rotation speed of $\approx 100$ km/s at $\approx 1$ Mpc from the axis (up to $4.2\sigma$ for dynamically cold filaments, viewed almost perpendicularly to their axis, and with the most massive halos at their end points). This corresponds to a local vorticity $v/r \approx 3\times 10^{-18}$ s$^{-1}$, of the same order as $H_0 \approx 2.3\times 10^{-18}$ s$^{-1}$, implying a rotation period of a few Hubble times. This local vorticity is, however, ten orders of magnitude larger than the upper limit on the global cosmic vorticity, $\sigma_V < 4.7\times 10^{-11}\, H_0 \approx 10^{-28}$ s$^{-1}$ \citep{SAADEH}, which quantitatively confirms that the filament rotation is a local dynamical phenomenon and cannot be a signature of global rotation. Similar effects were also found in N-body simulations of the standard cosmological model \citep{XIA}: vorticity is generated when structures form, with no need of a primordial rotation. Nevertheless, this result suggests a test closer to the original Gödel’s argument: to search for a coherent alignment of the sense of rotation of filaments -- or of galaxies -- over cosmological scales. Asymmetries in the distribution of the spin directions of spiral galaxies have been claimed (e.g. \citealt{SHAMIR}), but they are disputed (see the references therein).

\section{Vacuum, dark energy}
Thinking in terms of matter versus radiation, the vacuum is a serious challenge. Does it need some existence energy? If yes, then it would fall into the matter category. If it is pure temporal energy, then it would be radiation. If the space-time is quantized, then the vacuum could be the bound state of these quanta. But, again, it would fall into the matter category (in addition, one should also ask what might be the corresponding boson for this bound state). Also matter contains vacuum: a solid body is made of atoms forming a crystal lattice, but most of space of the macroscopic physical body is vacuum. If the vacuum is the bound state of quantized space-time, then would it be a compact aggregate or is there anything else?

If these considerations might seem specious and sophistic, I would like to underline the importance to understand what the stress-energy tensor in the Einstein's field equations is representing. As Einstein himself wrote \citep{EINSTEIN}: 

\begin{quote}
We make the distinction hereafter between gravitational field and matter in this way, that we denote everything but the gravitational field as matter. Our use of the word therefore includes not only matter in the ordinary sense, but the electromagnetic field as well. 
\end{quote}

Therefore, Einstein collected both matter and radiation under the word matter. If one thinks in terms of energy, this is a clear and justified choice, at least at a first order, having not included the energy of the gravitational field. If one thinks in terms of mass, then why putting together matter (mass) and radiation (massless)? 

Einstein then added -- and later deleted -- to his field equations the cosmological constant $\Lambda$, which is today supported by the observations of type Ia supernovae \citep{RIESS98,PERLMUTTER}, of the CMB, and of the baryon acoustic oscillations \citep{PLANCK}. The common interpretation of $\Lambda$ is that it represents a dark energy, which makes the Universe to accelerate its expansion. However, the tension in the measurements of the Hubble constant \citep{RIESS} is a different issue: it is a discrepancy that cannot be resolved within the standard $\Lambda$CDM model and is not evidence for $\Lambda$ itself, but may instead require new physics in the early universe, possibly in the form of a transient ``early dark energy'' component that contributes $\sim 10\%$ of the total energy density briefly before recombination \citep{KAMIONKOWSKI}. 

According to some authors $\Lambda$ might be a zero-point curvature, an intrinsic property of space-time \citep{BIANCHI1,BIANCHI2}, others suggested the existence of an exotic fluid, which generates negative pressure when interacting with baryonic matter \citep{CADONI18A,CADONI18B}, but most researchers explain the constant as the effect of a dark energy, identified as the quantum vacuum energy. However, the calculated values disagree by many orders of magnitude and many hypotheses have been suggested (e.g. \citealt{WEINBERG,RUGH,BLINNIKOV}). It is worth noting that the same mechanisms that explain the origin of mass make the problem worse: the condensates of quantum chromodynamics and of the Higgs field imply vacuum energy densities of the order of the fourth power of their energy scales, which already for the former ($\approx 200$~MeV) is $\sim 10^{44}$ times the observed value \citep{MARTIN}. The mass density of empty space is indeed listed by \cite{WILCZEK02} among the open problems on the origin of mass.

Also in this case, some notes are worth writing. Current quantum gravity theories are linked to the Planck units, which are physical quantities as functions of universal constants only (see \citealt{HOSS} for a detailed review). Current theories should break down at the Planck energy $E_{\rm P}$, where space-time should be quantized:

\begin{equation}
E_{\rm P}= \sqrt{\frac{\hbar c^5}{G}} \approx 1.9\times 10^{9}~\mathrm{J} = 1.2\times 10^{19}~\mathrm{GeV}
\end{equation}

which is equivalent to a mass of $\approx 2.2\times 10^{-8}$~kg. It is the mass for which the Compton wavelength equals the gravitational radius. According to \cite{WILCZEK02}, in these units the question is not why gravity is so feeble -- between two protons, it is $\sim 10^{36}$ times weaker than the electric repulsion -- but why the proton is so light, $Gm_{\rm p}^2/\hbar c\approx 6\times 10^{-39}$, and the answer is in the logarithmic running of the coupling of quantum chromodynamics. When calculating the quantum vacuum energy, one takes into account the zero-point energies of the electromagnetic field (fermion fields contribute with opposite sign and would result in a vacuum with positive pressure) with a cut-off at Planck scale (e.g. \citealt{AMENDOLA}). Just to write down some order-of-magnitude calculation, the zero-point energy density $\rho_{\rm vac}$ in the frequency range $[\omega_1,\omega_2]$ is \citep{MILONNI}:

\begin{equation}
\rho_{\rm vac} = \int_{\omega_1}^{\omega_2} \frac{\hbar \omega^3}{2\pi^2 c^3} d\omega = \frac{\hbar}{8\pi^2 c^3}(\omega_2^4-\omega_1^4)=\frac{h\pi}{c^3}(\nu_2^4-\nu_1^4)
\label{eq:vac}
\end{equation}

where $\omega=2\pi\nu$. The Planck angular frequency is:

\begin{equation}
\omega_{\rm P} = \frac{E_{\rm P}}{\hbar} = t_{\rm P}^{-1} = \sqrt{\frac{c^5}{\hbar G}} \approx 1.9\times 10^{43}~\mathrm{rad/s}
\end{equation}

which corresponds to $\nu_{\rm P}=\omega_{\rm P}/2\pi\approx 3.0\times 10^{42}$~Hz(\footnote{In the previous versions of this essay, $\omega_{\rm P}$ was inserted in Eq.~(\ref{eq:vac}) as a frequency $\nu$, thus overestimating $\rho_{\rm vac}$ by a factor $(2\pi)^4\approx 1.6\times 10^{3}$. This does not change the argument.}). Therefore, by taking $\nu_{1}=0$ and $\nu_{2}=\nu_{\rm P}$, it results $\rho_{\rm vac}\approx 6\times 10^{112}$~erg/cm$^3$. This value must be compared with that observed on cosmological scale: $\rho_{\Lambda}\approx 5\times 10^{-9}$~erg/cm$^3$ \citep{PLANCK}. The difference is crystal clear. Pauli, in the 1920s, had already noted that if the zero-point energy of the electromagnetic field were cut off at the classical electron radius, the radius of curvature of the universe would be about 31~km -- not even reaching the Moon \citep{RUGH}.

If one forgets about the quantization of space-time and simply adopts the observed value $\rho_{\Lambda}$ to set a constraint on the cut-off frequency, then rearranging Eq.~(\ref{eq:vac}) and setting $\rho_{\rm vac}=\rho_{\Lambda}$, one obtains:

\begin{equation}
\nu_{\rm cut} = \sqrt[4]{\frac{\rho_{\Lambda}c^3}{h\pi}}\approx 1.6\times 10^{12}~\mathrm{Hz}
\end{equation}

Interestingly, the present day peak frequency of the spectral radiance of the Cosmic Microwave Background (CMB) is $\approx 2.2\times 10^{11}$~Hz, so the two values are consistent within one order of magnitude. In the previous versions of this essay, I speculated that the vacuum, as zero-point energy of the electromagnetic field, might be generated at the epoch of the photon decoupling, so that the CMB would be not only the last scattering surface, but also a sort of dark energy horizon. A closer look shows that this is not tenable, for the following reasons.

First, the comparison mixes two different epochs. The above cited peak frequency of the CMB is the present-day value, for a temperature $T_0\approx 2.725$~K. At the epoch of decoupling ($z\approx 1100$), the temperature was $\approx 3000$~K and the peak frequency was $(1.8-3.1)\times 10^{14}$~Hz, two orders of magnitude greater than $\nu_{\rm cut}$. Moreover, to mimic a cosmological constant, $\nu_{\rm cut}$ must be constant in time, while the frequency of the CMB scales as $(1+z)$. Therefore, the two frequencies are close to each other only today.

Second, a physical cutoff at $\nu_{\rm cut}$ is excluded by laboratory experiments. The Casimir force was measured between surfaces separated by distances from a fraction of a micron to a few microns \citep{LAMOREAUX}, where the relevant field modes have wavelengths comparable with the separation, much shorter than $c/\nu_{\rm cut}\approx 190~\mu$m. The Lamb shift and the spontaneous emission involve vacuum fluctuations at much higher frequencies \citep{MILONNI}. Therefore, the zero-point fluctuations above $\nu_{\rm cut}$ do exist and have measurable electromagnetic effects. The open question is then why they do not contribute to the dark energy as naively expected, which is the cosmological constant problem itself.

Third, the opacity of the universe before decoupling does not imply the absence, or the suppression, of the modes of the electromagnetic field. The primordial plasma was filled with radiation in thermal equilibrium, which is what is observed today as CMB. The opacity was due to the Thomson scattering, whose cross section does not depend on frequency, and thus cannot select any frequency scale. A plasma does modify the field modes, but only below the plasma frequency, which at decoupling was $\lesssim 2\times 10^{5}$~Hz (for an electron density $\lesssim 3\times 10^{2}$~cm$^{-3}$), many orders of magnitude below $\nu_{\rm cut}$.

Fourth, a sharp cutoff in frequency is not Lorentz invariant. The vacuum regularized in this way has pressure $p=\rho/3$, like radiation, instead of $p=-\rho$, as required by a cosmological constant. A regularization that preserves the Lorentz invariance gives the correct equation of state, but the contribution of a field is then proportional to the fourth power of its mass, and vanishes for a massless field such as the electromagnetic one \citep{MARTIN}. Therefore, Eq.~(\ref{eq:vac}) with a cutoff can only be taken as an estimate of an energy scale ($h\nu_{\rm cut}\approx 7$~meV), not as a physical model of dark energy. In terms of a Compton frequency, $\nu_{\rm C}=mc^2/h$, this corresponds to a rest energy of the same order as the neutrino mass scales measured by oscillation experiments ($\sqrt{\Delta m^2}\approx 9$ and $\approx 50$~meV; \citealt{PDG}), a well-known coincidence for which no accepted explanation exists. It must be noted that a cutoff at the Compton frequency of the electron, the lightest charged particle and thus the only natural scale for the electromagnetic field, would give a value $\sim 10^{31}$ times greater than $\rho_{\Lambda}$.

Fifth, this proximity is not a new piece of information. Both energy densities scale with the fourth power of a frequency: Eq.~(\ref{eq:vac}) for the vacuum, and the Stefan-Boltzmann law for the CMB,

\begin{equation}
\rho_{\rm CMB} = \frac{\pi^2 (k_{\rm B}T_0)^4}{15\hbar^3 c^3}\approx 4.2\times 10^{-13}~\mathrm{erg/cm^3}
\end{equation}

where $k_{\rm B}$ is the Boltzmann constant. Then, $\rho_{\Lambda}/\rho_{\rm CMB}\sim 10^{4}$, whose fourth root is $\sim 10$. To state that the two frequencies are consistent within one order of magnitude is equivalent to state that today the energy density of radiation is about four orders of magnitude smaller than that of dark energy. This is one aspect of the well-known cosmic coincidence problem \citep{WEINBERG,AMENDOLA}: we live in an epoch when the energy densities of matter, radiation, and dark energy are not too far from each other, while they differ by many orders of magnitude for most of the history of the universe.

To summarize, the proximity of $\nu_{\rm cut}$ to the peak frequency of the CMB is a chance coincidence of the present epoch, and the speculations of the previous versions (the vacuum as zero-point energy of the decoupled radiation, and the CMB as dark energy horizon) must be dropped. I prefer to keep the calculation, together with the reasons why the coincidence is only apparent, rather than simply removing it.

\section{The Universe as a Kerr-deSitter black hole}
In the previous versions of this essay, I also speculated that the CMB would not only be the last scattering surface, but also the dark energy horizon. The Universe may be thought of as a Kerr-de Sitter black hole. The comoving radial distance of the last scattering surface ($\approx 4.3\times 10^{28}$~cm, by adopting the cosmological parameters measured by the Planck Collaboration, \citealt{PLANCK}) was taken as the additional horizon generated by $\Lambda$, obtaining $\Lambda\sim 10^{-57}$~cm$^{-2}$ and an almost maximally rotating universe. I withdraw also this argument. That distance is the present-day comoving one (the proper distance at $z\approx 1100$ was $\approx 4\times 10^{25}$~cm), and it is of the same order of the present Hubble radius ($c/H_0\approx 1.4\times 10^{28}$~cm) and of the de Sitter radius ($\sqrt{3/\Lambda}\approx 1.7\times 10^{28}$~cm). Any length $R$ of this order gives $\Lambda\sim R^{-2}\sim 10^{-57}-10^{-56}$~cm$^{-2}$, so that to recover the order of magnitude of $\Lambda$ is not an independent confirmation: it is again the cosmic coincidence. Moreover, an almost maximally rotating universe is in conflict with the upper limits on the cosmic vorticity cited in the previous section \citep{SAADEH}.

As a side note on horizons and $\Lambda$, one might think of astrophysical black holes. The generally adopted Kerr metric can be transformed into a Kerr-deSitter space-time by adding a dark energy constant $\Lambda$ (\citealt{AKCAY}, the two metrics are coincident for $\Lambda = 0$). The presence of dark energy in the latter metric implies one more horizon in addition to the gravitational radius. The problem is that the current observed value of $\Lambda\sim 10^{-56}$~cm$^{-2}$ is too small to generate the additional horizon on local scale. By using the example proposed by \cite{AKCAY} and one of the most massive black hole known (S5~$0014+813$, $z=3.366$, $M\approx 4\times 10^{10}M_{\odot}$, \citealt{GHISELLINI}), it results that to obtain an observable dark energy horizon of $r_{\rm deh}\approx 4\times r_{\rm g}\approx 2.4\times 10^{16}$~cm, it would be necessary $\Lambda\approx 3\times 10^{-33}$~cm$^{-2}$, about 23 orders of magnitude greater than the known cosmological $\Lambda$. In addition, decades of observations have never shown significant and anomalous deviations from the Kerr metric (e.g. \citealt{FOSCHINI2025}). 

\section{Energy}
Focusing on time, and hence on energy, opens interesting research directions. As noted in my previous work \citep{FOSCHINI}, matter needs an existence energy, the minimum to exist at rest, just to travel in time. The key experiment is to measure the lifetime of protons and electrons, which is currently just a lower bound (e.g. \citealt{AGOSTINI}). A measured value will also allow us to set a constraint on the lifetime of the universe, if any. Further questions might be how and where this energy is dissipated: is it simply lost? Or is it converted? In the first case, this clearly violates the principle of energy conservation, which was already challenged on cosmological scales, given the arrow of the expansion of the universe (see \citealt{VELTEN} for a recent review). If the existence energy is converted, then the energy conservation might be saved, but what other form of energy is the result of this conversion? An explicit, although classical, model of a massive object slowly losing its rest energy is again the geon, whose trapped radiation leaks away with a calculable lifetime \citep{WHEELER}.

\section{Conclusions}
I wanted to show how new research directions are emerging just by abandoning the mass-particle paradigm. Mass without mass is an old idea \citep{WHEELER,WILCZEK02,WILCZEK12}: mass is not removed, but it loses its status of primitive concept, becoming the energy of fields in their rest frame. Instead of searching for new exotic particles or unnecessary quantizations, one might explore new space-time metrics, new types of energy, changes in the rhythm of time. I think that the current mixture of classical and frontier concepts, which is keeping us bound to the concept of mass, might be lethal for physics in the long term, while focusing on time and energy opens new, and interesting research lines. 

The link between the vacuum energy cutoff frequency and the cosmic microwave background -- proposed in earlier versions of this essay -- along with the speculations derived from it, have been withdrawn in this version, as I was mistaken. My error was to compare measurements from two different epochs. I have chosen to update this essay -- retaining some of the more philosophical reflections -- rather than withdrawing it entirely.

\vspace{1cm}
\noindent \textbf{Acknowledgements:} I would like to warmly thank Roberto Casadio and Alberto Vecchiato for their comments and advices. Thanks also to Michelle Galloway for pointing out an error in the citation of the XENON experiment results, and to Marcel-Marie LeBel for recommending the important Unruh's essay.

\end{document}